\documentstyle[12pt]{article}
\def\be{\begin{equation}}
\def\ee{\end{equation}}

\def\f{\frac}

\def\bea{\begin{eqnarray}}
\def\eea{\end{eqnarray}}

\def\ll{\langle\langle}
\def\rr{\rangle\rangle}
\def\rot{\mathop{\mathrm{rot}}}
\def\grad{\mathop{\mathrm{grad}}}
\def\div{\mathop{\mathrm{div}}}

\def\ben{\begin{displaymath}}
\def\een{\end{displaymath}}
\def\ba{\begin{array}{c}}
\def\bal{\begin{array}{l}}
\def\ea{\end{array}}
\def\p{\partial}

\begin{document}


\vspace{1.5cm}

 \begin{center}{\Large \bf
  Relativistic vector bosons and PT-symmetry

  }\end{center}

\vspace{10mm}

 \begin{center}

{\bf Jaroslav Smejkal}

\vspace{3mm}

 \'{U}stav technick{\'{e}} a experimentaln{\'{\i}} fyziky  \v{C}VUT

 Horsk\'{a} 3a/22,
128 000  Praha 2\,--\,Nov{\'{e}}
 M{\v{e}}sto

Czech Republic

{e-mail: smejkal@ujf.cas.cz}

\vspace{5mm}

 {\bf V\'{\i}t Jakubsk\'{y}}

 \vspace{3mm}

 \'{U}stav jadern\'e fyziky AV \v{C}R

250 68 \v{R}e\v{z}

Czech Republic

{e-mail: jakub@ujf.cas.cz}

\vspace{5mm}

and

\vspace{3mm}

 {\bf Miloslav Znojil}

 \vspace{3mm}
 \'{U}stav jadern\'e fyziky AV \v{C}R

250 68 \v{R}e\v{z}

Czech Republic

{e-mail: znojil@ujf.cas.cz}


\vspace{5mm}


\end{center}

\vspace{5mm}

\section*{Abstract}

Relativistic massive bosons with spin one are considered in
several quantization schemes. In all of them the system is shown
described by a non-Hermitian Hamiltonian $H \neq H^\dagger$ and
helicity operator $\Lambda$. Constructively we show that in all of
the contemplated schemes both these operators $H$ and $\Lambda$
prove simultaneously ${\cal PT}-$symmetric, i.e., pseudo-Hermitian
with respect to a certain not too complicated indefinite
pseudo-metric operator ${\cal P}={\cal P}^\dagger$.

\subsection*{Keywords}

Relativistic vector bosons, quantization, Dirac theory

\subsection*{PACS}

03.50.Kk;  03.65.Ca;   03.65.Pm;

\newpage

\section{Introduction}

The majority of textbooks on Quantum Mechanics illustrates the
performance of the formalism via a non-relativistic point particle
of mass $m$ which moves in a one-dimensional potential $V(x)$. On
the basis of the principle of correspondence the time evolution of
such a system is quite often easily described as generated by the
ordinary differential self-adjoint Hamiltonian operator
$H=\hat{p}^2+V(x)$ where $\hat{p}^2 = - d^2/dx^2$ in units $\hbar
= 2m = 1$.
Much less attention is usually being paid to the possibilities of
the transition to relativistic kinematics \cite{Greiner}, i.e., to
the Klein-Gordon equation for the spinless bosons, to the Dirac
equation for the simplest fermions, to the Maxwell equations for
the massless photons, to the Proca equation for the massive bosons
with spin one \cite{paraKGDirac} etc.

It is well known that the correct description of relativistic
systems requires the use of the full-fledged formalism of
relativistic Quantum Field Theory. Another, ``hidden" reason of
the overall reluctance of working with relativistic kinematics
even in an approximative regime of Quantum Mechanics is that it
often requires a transition to non-self-adjoint generators of the
time evolution, i.e., to the Feshbach-Villars Hamiltonian operator
$H^{\rm (FV)}\neq \left (H^{\rm (FV)}\right )^\dagger$ in the
spinless case \cite{FV} etc. Apparently, the naive use of the
similar Hamiltonians may lead to inconsistencies. For example,
negative probabilities of the localization are traditionally being
mentioned as occurring for the relativistic spinless boson at a
point $x$ \cite{Constantinescu}, etc.

The formalism based on the FV-like Hamiltonians $H^{\rm (FV)}$
need not necessarily be mathematically inconsistent
\cite{Ali,zno}. In fact, the use of $H=H^{\rm (FV)}$ may be
re-interpreted as one of the most characteristic applications of
the so called PT-symmetric version of Quantum Mechanics as
originally proposed by Bender and Boettcher \cite{BB}. Thanks to a
concentrated effort and subsequent debate (cf. its sample in
\cite{ostatni}--\cite{last}) it became clear that even the
manifestly non-Hermitian PT-symmetric operators of observables
${\cal O}$ (characterized, for our present purposes, by their
property
 \be
 {\cal O}^\dagger={\cal P}\,{\cal O}\,{\cal P}^{-1} \neq {\cal O}
 \label{PTS}
 \ee
with a suitable ``indefinite metric" operator ${\cal P}$ or with
some of its non-metric alternatives \cite{pneq}) may fit in the
overall scheme of Quantum Mechanics.

On this formal background and in the language set already by
Scholtz et al in early nineties \cite{Geyer} it has been
recognized that in the relativistic spinless case the Hamiltonian
$H^{\rm (FV)}$ may be treated as self-adjoint with respect to some
less standard inner product in Hilbert space. In the other words,
the use of the non-Hermitian observables may often be made fully
acceptable via an introduction of a certain nontrivial, positive
definite ``physical" metric operator $\Theta \neq I$ \cite{BBJo}
-- \cite{Geyer}.

All the ``physical" choices of $\Theta$ {\em must} make {\em all}
the underlying observables (say, ${\cal O}= H^{\rm (FV)}$ etc)
self-adjoint in some new, $\Theta-$dependent Hilbert space,
 \be
 {\cal O}^\dagger=\Theta\,{\cal O}\,\Theta^{-1} \neq {\cal O},
 \ \ \ \ \ \ \ \Theta = \Theta^\dagger > 0.
 \label{QH}
 \ee
In this sense, no modification of the postulates of Quantum
Mechanics is needed. Nevertheless, in order to avoid confusion the
authors of the review \cite{Geyer} recommended that one should
speak about quasi-Hermiticity of ${\cal O}$ whenever $\Theta \neq
I$ in (\ref{QH}).

For the relativistic spinless system the ``weakened" Hermiticity
property (\ref{QH}) offered a natural explanation of the
negative-probability Klein-Gordon puzzle \cite{zno}. In what
follows we intend to discuss some aspects of the transition to the
``next" case of the {massive field with spin one}. The main
emphasis of our paper will be put on the existence of the ${\cal
PT}-$symmetry features of this system and on the emergence of
their specific ambiguities which arise in connection with the
presence of a nontrivial spin $s > 0$ which mediates a coupling of
channels \cite{coupledchannels}.

Our present results will be separated into their classical part
(section \ref{dva}), quantization part (section \ref{tri}) and
${\cal PT}-$symmetry-related part (section \ref{styria}). Our
observations starting from the classical case may be perceived as
separating the Lagrangian and Hamiltonian formulations, with the
core of our attention concentrated upon the latter language. Under
both the classical and relativistic kinematics our study will be
sub-separated into the economical ``non-Dirac approach" without
any constraints (cf. subsections \ref{eko} and \ref{alfa} or
\ref{sixD}, respectively) and the more common ``Dirac approach"
using constraints (cf. the respective subsections \ref{diro} and
\ref{beta} or \ref{eightD}).

In the latter, Dirac-theory context there still exists certain
freedom in the choice of the method of the quantization. Three
possibilities, viz., ``version A" (cf. \ref{A} or \ref{beta}),
``version B" (cf. \ref{B} or \ref{gama}) and ``version C" (cf.
\ref{C} or \ref{delta}) are studied in more detail in what
follows.

As long as the key purpose of our study lies in an extension of
the concept of ${\cal PT}-$symmetry (\ref{PTS}) to the equations
controlling the motion of the relativistic vector bosons, section
\ref{styria} represents in fact a climax of our considerations. A
number of explicit formulae is derived and displayed there
including the matrix forms of the Hamiltonian as well as of the
related pseudo-metric operator ${\cal P}$.

Our last section \ref{sest} brings a brief summary of our results
while Appendix~A adds a brief discussion of a few parallels
between our present (massive) model and its more common massless
(i.e., electromagnetic) counterpart.

\section{Bosons with spin one \label{dva}}

Our project has been inspired by the description of the massive
vector boson system by Taketani, Sakata and Tamm \cite{ST} as
summarized by Nikitin et al \cite{Nikitin}. Our sample of relevant
references shouldn't also omit Kemmer \cite{Kemmer} (in the
context of the so-called Kemmer-Duffin-Petiau equation) or Shay
and Good and Weinberg \cite{many} (in a slightly different
context) and many others, with a nice summary of the field offered
by Labent\'e \cite{Labente}.

\subsection{Lagrangians}

Let us start now our present discussion from the classical case
and assume that the dynamics of the classical complex vector field
${A}$ of mass $m$ is given by its Lagrangian density
 \ben
{\cal{L}}=m^2\bar{{A}}_{\mu}{A}^{\mu}- {\textstyle \frac{1}{2}}
 \left(\partial_{\mu}\bar{{A}}_{\nu}
 -\partial_{\nu}\bar{{A}}_{\mu}\right)
 \left(\partial^{\mu}{{A}}^{\nu}-\partial^{\nu}{{A}}^{\mu}\right).
 \een
The related action ${\cal{S}} \!=\! \int {\cal{L}} \:\!  d^4x$ has
to be stationary, $\delta {\cal{S}} \!=\! 0$. With the current
abbreviation $ \Box=-g^{\mu\nu}\partial_{\mu}\partial_{\nu}$ where
$ g^{\mu\nu} = {\rm diag}(1,-1,-1,-1)$ (and in units such that
$c=1$ and $x_0=t$) this gives the usual Euler-Lagrange equation
 \be
  \partial_{\mu}F^{\mu\nu}+m^2A^{\nu}=0,\ \ \ \ \
 F^{\mu\nu}=\partial^{\mu}A^{\nu}-\partial^{\nu}A^{\mu}
 \label{1}
 \ee
plus its complex conjugate. Its differentiation and summation
reveals that the individual components of the field have to
fulfill Klein-Gordon equation subject to the free
Klein-Gordon-type constraint,
 \be
 \left(m^2-\Box\right){A}_{\mu}=0,\ \ \ \ \ \ \  \
 \partial_{\nu}{{A}}^{\nu}=0.
  \label{klien}
  \ee
Consequently, only three of the four components of the field are
linearly independent. The  equation  itself is obviously satisfied
by the wave solutions distinguished by the sign $\alpha=\pm$ of
the energy,
 \be
  A_{\alpha}^{\mu}(x)=
 N_{\alpha}(p)u^{\mu}_{\alpha}(p) {\rm e}^{-i{p}{x}},
 \ \ \ \ \
 \ \ \ p_0=\alpha \omega,
 \ \ \ \ \ \omega=|\sqrt{{\bf p}^2+m^2}|.
 \label{3}
 \ee
An additional constraint acquires the form
$
 p\,u_{\alpha}(p) \!=\! 0 $.
For the sake of definiteness the three linearly independent
four-vectors $u_{\alpha}(p)$ may be chosen in the following form
 \be
 u_{\alpha,1}=(0,{\bf u}_{\alpha,1})^T,\ \ \ u_{\alpha,2}=(0,
 {\bf u}_{\alpha,2})^T
,
 \ \ \
 u_{\alpha,3}=\left(\alpha\frac{|{\bf p}|}
 {\omega},\frac{{\bf p}}{|{\bf p}|}\right)^T.
 \label{5}
 \ee
The space-like components (i.e., polarization vectors) in the
first two items are perpendicular to ${\bf p}$ while the third one
is chosen parallel to ${\bf p}$. We may fix
 \be
 {\bf u}_{\alpha,1}({\bf p})=\frac{(0,-p_3,p_2)^T}{\sqrt{p_3^2+p_2^2}},\
 \ {\bf u}_{\alpha,2}({\bf p})
 =\frac{(p_2^2+p_3^2,-p_1p_2,-p_1p_3)^T}{\sqrt{(p_3^2+p_2^2)^2
 +p_1^2p_2^2+p_1^2p_3^2}}.
 \label{7}
 \ee
Alternatively,  the operator of helicity
 \be
 \hat{h}({\bf p})=\frac{{\bf p}{\bf S}}{|{\bf p}|}={\bf n}{\bf S}=
 i\left(\begin{array}{ccc}0&-n_3&n_2\\n_3&0&-n_1\\-n_2&n_1&0
 \end{array}\right)
 \label{8}
 \ee
becomes defined in terms of the three generators of the
three-dimensional representation of the rotation group,
 \ben
 S_1=\left(\begin{array}{ccc}0&0&0\\0&0&-i\\0&i&0\end{array}\right),\ \
 S_2=\left(\begin{array}{ccc}0&0&i\\0&0&0\\-i&0&0
 \end{array}\right),\ \
 S_3=\left(\begin{array}{ccc}0&-i&0\\i&0&0\\0&0&0\end{array}
 \right)\,.
  \een
Its action on the vectors (\ref{5}), [$\hat{h}{\bf
u}_{\alpha,1}=i{\bf u}_{\alpha,2}$, $\hat{h}{\bf
u}_{\alpha,2}=-i{\bf u}_{\alpha,1}$ and $\hat{h}{\bf
u}_{\alpha,3}=0$ (with subscript $\alpha=\pm$)] prompts an
introduction of the helicity-numbered eigenvectors ${\bf
u}_{\alpha}(p,\pm 1)$,
 \be
 \hat{h}{\bf u}_{\alpha}(p,\pm 1)=\pm{\bf u}_{\alpha}(p,\pm 1),\ \
 \hat{h}{\bf u}_{\alpha}(p,0)=0,
 \label{10}
 \ee
with an explicit representation $
 {u}_{\pm}(p,+1)=({u}_{\pm,1}+i{u}_{\pm,2})/{\sqrt{2}}$,
 ${u}_{\pm}(p,-1)=({u}_{\pm,1}-i{u}_{\pm,2})/{\sqrt{2}}$ and
 ${u}_{\pm}(p,0)={u}_{\pm,3}(p)$.
Combining (\ref{3}), (\ref{5}) and (\ref{7})  we get
 \be
 A^{\mu}_{\pm,h}=N_{\pm,h}u_{\pm}^{\mu}(p,h)
 {\rm e}^{i{\bf p}{\bf x}\mp i\omega t}
 \label{11}
 \ee
i.e., waves traveling at a fixed energy and helicity.

\subsection{Hamiltonians}

There exist, basically, two ways of the derivation of the
description of the time evolution of the vector boson fields from
the first principles. In one of them one eliminates all the
(redundant) degrees of freedom from the very beginning. We shall
call this approach ``economical" because the number of its field
``parameters" is, in  some sense, minimal and equal to the number
of the degrees of freedom of the field (cf. also subsection
\ref{eko} for more details).

An alternative recipe (cf. the subsequent subsection \ref{diro})
carries the name of Dirac \cite{Duirac}. It employs the language
of constraints and its key merit may be seen in its more formal,
``algorithmic" character. It offers a virtually unambiguous recipe
for a systematic treatment of the field in question. Both its
physical and non-physical components are treated ``on an equal
footing".

The key purpose of our present paper lies in the discussion of the
latter two approaches. We intend to illustrate (by construction!)
that both of them remain equivalent on the classical level while a
number of open question can arise during their quantization.

Before we fully concentrate upon the questions of specific
differences between quantization recipes for vector bosons (in
which the classical fields should be replaced by operators) we
intend to emphasize that several important technicalities must
already be re-analyzed in a preparatory step, on the classical
level. On this level the first relevant aspect of the problem may
be determined immediately as lying in the existence of the
``economical" possibility which looks nonstandard and which seems
to be in an apparent conflict with the more modern formulations of
the problem.

In the modern (usually called Dirac's) theory the quantization is
treated with the BRST and Batalin-Vilkovisky methods
\cite{Batalin}. For a more detailed elucidation of the topic the
reader might consult the related literature where the quantization
of the relativistic (free) particles is reminiscent of Feynman's
fifth time formalism and where the theory involves the
quantization of the reparametrization-invariant models on the
world line, etc.

The latter approach leads to the Klein-Gordon equation emerging as
a constraint (analogous to the Gauss law in QED) and it opens new
perspectives, i.a., in supersymmetry. At the same time it does not
represent the only possibility and it may often be complemented by
the less subtle approach where the Klein-Gordon equation results
from the ``economical" process where one {\em starts} from an
elimination of {\em all} the redundant degrees of freedom. Thus,
once we succeed in a minimization of the number of the fields we
are not forced to impose any auxiliary constraints. We shall add
more details in paragraph \ref{eko}.

One could stress that the minimization strategy circumvents (or at
least clarifies) certain shortcomings of the use of the Dirac's
eight-dimensional Hamiltonians by their replacement by their
compact, six-dimensional ``economical" alternatives. {\it Vice
versa}, the specific merits of the eight-dimensional formulae will
be recollected in paragraph \ref{diro}. We shall clarify there the
distinction between several alternative ways of the detailed
application of the Dirac's recipe. We shall distinguish between
certain ``approach A" (cf. paragraph \ref{A} for more details),
``approach B" (cf. paragraph \ref{B}) and ``approach C" (cf.
paragraph \ref{C}). We believe that in this way an enhanced
transparency will be assigned not only to the essence of the
(second-class) constraints but also to the subsequent questions of
the quantization.

\subsection{``Economical"  formulation \label{eko}}

Using the standard recipe and taking ${A}_{\mu}$ and
$\bar{{A}}_{\mu}$ as generalized coordinates we get the
generalized momenta using the standard recipe,
 \ben
 \Pi_m=\frac{\partial \cal{L}}{\partial(\partial_0 {A}_m)}
 =\partial_0\bar{{A}}_m-\partial_m \bar{{A}}_0,\ \ \ \ \
   \Pi_0=\frac{\partial \cal{L}}{\partial(\partial_0 {A}_0)}=0,
   \een
   \be
 \bar{\Pi}_m=\frac{\partial \cal{L}}
 {\partial(\partial_0 \bar{{A}}_m)
 }=\partial_0{{A}}_m-\partial_m {A}_0,\ \ \ \ \
   \bar{\Pi}_0=\frac{\partial \cal{L}}
   {\partial(\partial_0 \bar{{A}}_0)
   }=0.
   \label{eq:dva}
   \ee
We see that the time-coordinate momenta $\Pi_{0}$ and
$\bar{\Pi}_{0}$ can be omitted whereas their respective field
conjugates can be eliminated using the equations of
motion~(\ref{1}),
 \be A_{0} =
\frac{1}{m^{2}} \, ( \partial_{m} \bar{\Pi}_{m} ), \ \ \ \ \
\bar{A}_{0} = \frac{1}{m^{2}} \, ( \partial_{m} \Pi_{m} ).
\label{eq:jedna}
 \ee
After a straightforward calculation the density of the Hamiltonian
of our classical vector field can be found,
 \ben
 {\cal{H}}=\Pi_m(\partial_0 A_m)+\bar{\Pi}_m(\partial_0\bar{{A}}_m)
 -{\cal{L}} =
 \een
 \ben
 =\bar{\Pi}_m\Pi_m
 +\frac{1}{m^2}(\partial_m\bar{\Pi}_m)(\partial_n{\Pi}_n)
 +m^2\bar{{A}}_m{A}_m+
 (\partial_m\bar{{A}}_n)(\partial_m{{A}}_n)-(\partial_m\bar{{A}}_m)
 (\partial_n{{A}}_n). \nonumber
 \een
Classically, Euler-Lagrange equations (\ref{klien}) correspond to
the dynamical equations
 \be
 \dot{{A}}_m=\{H,{A}_m\}_{\rm P},\ \ \dot{\Pi}_m=\{H,\Pi_m\}_{\rm P}
 \label{pois}
 \ee
where $\{.,.\}_{\rm P}$ denotes the Poisson bracket and where $H
\!=\! \int {\cal{H}} \;\! d{\bf x}$.

\subsection{Dirac theory using constraints \label{diro}}

We already mentioned that there are several possibilities of the
strictly formal transition from Lagrangians to Hamiltonians.
Sometimes, one has to interpret the impossibility of the
elimination of the generalized velocities as an emergence of
certain phase-space constraints (to be denoted as $\theta$). They
are called ``primary constraints" and will be marked by the
superscript $^{\rm (I)}$. We shall have to deal with a pair of
them in our vector-boson model.

In addition, the natural requirement of the time-independence of
the constraints $\theta^{\rm (I)}$ leads, in principle, to the
emergence of the further series of constraints denoted as
$\theta^{\rm (II)}$. They are called ``secondary" and we will have
a pair of them in what follows.

In version ``A" of the formalism (cf. paragraph \ref{A} below),
the Hamiltonians will be constructed, using the technique of the
Lagrange multipliers, in a way which employs the primary
constraints $\theta^{\rm (I)}$ only. In version ``B" of paragraph
\ref{B}, the knowledge of {\em all} the constraints $\theta^{\rm
(I)}$ and $\theta^{\rm (II)}$ will be employed giving an
alternative family of the Hamiltonian functions. Finally, the
approach ``C" of paragraph \ref{C} will re-define the brackets and
enable us to prepare and perform the quantization in an internally
most consistent and compact manner.

\subsubsection{Approach A \label{A}}

In the approach ``A" one has to apply the Lagrange multipliers in
order to respect the existence of the two primary constraints.
These constraints read
 \be
\theta^{\rm (I)} = \Pi_{0} = 0, \ \ \ \ \ \bar{\theta}^{\rm
(I)}=\bar{\Pi}_{0} = 0 \label{eq:tri}
 \ee
and emerge in the process of the transformation of the Lagrangian
into the Hamiltonian (cf.\ also the formulae of
eq.~(\ref{eq:dva})). An ``extended" Hamiltonian ${\cal H}^{\rm
(A)}$ will be then sought as containing a pair of the Lagrange
multipliers $\lambda^{\rm (I)}$ and $\bar{\lambda}^{\rm (I)}$,
 \be
{\cal H}^{\rm (A)} = {\cal H}^{0} +
  \lambda^{\rm (I)} \theta^{\rm (I)} +
  \bar{\lambda}^{\rm (I)} \bar{\theta}^{\rm (I)}.
\label{eq:ctyri}
 \ee
The symbol ${\cal H}^{0}$ stands for the standard, non-extended
Hamiltonian function,
\begin{eqnarray}
{\cal H}^{0}
  &=&
  \Pi_m(\partial_0 A_m)+\bar{\Pi}_m(\partial_0\bar{{A}}_m)
 - {\cal{L}} =
 \label{eq:pet} \\
  &=&
  \bar{\Pi}_m\Pi_m
 +
    m^2\bar{{A}}_m{A}_m -
    m^2 \bar{A}_{0} {A}_{0} +
    (\partial_m\bar{{A}}_n)(\partial_m{{A}}_n)-
    \nonumber \\
  && -(\partial_m\bar{{A}}_m)
    (\partial_n{{A}}_n) + \Pi_{m} (\partial_{m} A_{0}) +
    \bar{\Pi}_{m} (\partial_{m} \bar{A}_{0}).
    \nonumber
\end{eqnarray}
In the next step one demands the time independence of the primary
constraints,
 \be
  \{ {{H}^{\rm (A)}} , \theta^{\rm (I)} \}_{\rm P} = 0, \ \ \ \ \
  \{ {{H}^{\rm (A)}} , \bar{\theta}^{\rm (I)} \}_{\rm P} = 0,
\label{eq:devet}
 \ee
where
 \be
  {H}^{\rm (A)} = \int d \vec{x} \, {\cal H}^{\rm (A)}.
\label{eq:deset}
 \ee
In our present model such a requirement gives rise to the mere
doublet of secondary constraints,
 \be
 \theta^{\rm (II)} = - m^{2} \bar{A}_{0} -
 ( \partial_{m} \bar{\Pi}_{m} ) =
  0, \ \ \ \ \
 \theta^{\rm (II)} = - m^{2} {A}_{0} -
 ( \partial_{m} \bar{\Pi}_{m} ).
 \label{eq:sest}
 \ee
In the subsequent step we have to demand the time independence of
the secondary constraints, too. In a way paralleling
eq.~(\ref{eq:devet}) we obtain the Lagrange multipliers
$\lambda^{\rm (I)}$ and $\bar{\lambda}^{\rm (I)}$ as a solution.
In this way we arrive at the final Hamiltonian
 \be
{\cal H}^{\rm (A)} = {\cal H}^{0} +
  \Pi_{0} (\partial_{m} A_{m}) + \bar{\Pi}_{0} (\partial_{m}
   \bar{A}_{m}).
\label{eq:sedm}
 \ee
The process of construction guarantees that all the equations of
motion respect all the primary {\em and} secondary constraints.
The same information is of course carried also by equations
(\ref{eq:jedna}) which were employed in the ``economical" approach
of paragraph \ref{eko}.

\subsubsection{Version B \label{B}}

In the previous approach A, the secondary constraints were clearly
discriminated in comparison with the primary ones. Fortunately, in
a way outlined, e.g., in the popular textbook \cite{GT} we can
also introduce all the constraints in the Hamiltonian {\em
simultaneously}, regardless of their origin.

First steps of such a slightly more sophisticated approach to the
construction of Hamiltonians are exactly the same as in the
approach A. The differences appear after one obtains {\em all} the
secondary constraints. Having collected all the constraints
$\theta^{\rm (I)}$ and $\theta^{\rm (II)}$ one decides to employ
them {\em all} in the recipe using Lagrange multipliers. This
means that our new postulated Hamiltonian ${\cal H}^{\rm (B)}$
will read
 \be
{\cal H}^{\rm (B)} = {\cal H}^{0} +
  \lambda^{\rm (I)} \theta^{\rm (I)} +
    \bar{\lambda}^{\rm (I)} \bar{\theta}^{\rm (I)} +
  \lambda^{\rm (II)} \theta^{\rm (II)} +
    \bar{\lambda}^{\rm (II)} \bar{\theta}^{\rm (II)}
\label{eq:osm}
 \ee
where the symbols $\lambda^{\rm (II)}$ and $\bar{\lambda}^{\rm
(II)}$ denote two new Lagrange multipliers attached to the
secondary constraints. All the values of the lambda-multipliers
emerging in eq.~(\ref{eq:osm}) may be again extracted from the
requirement of the time independence,
 \be
  \{ {{H}^{\rm (B)}} , \theta \}_{\rm P} = 0, \ \ \ \ \
  \theta \in
    \{ {\theta}^{\rm (I)}, \bar{\theta}^{\rm (I)},
      {\theta}^{\rm (II)}, \bar{\theta}^{\rm (II)} \},
      \ \ \ \ \
  {H}^{\rm (B)} = \int d \vec{x} \, {\cal H}^{\rm (B)}.
\label{eq:jedenact} \label{eq:dvanact}
 \ee
The resulting Hamiltonian
 \ben
{\cal H}^{\rm (B)} = {\cal H}^{0} +
  \Pi_{0} (\partial_{m} A_{m}) + \bar{\Pi}_{0} (\partial_{m}
   \bar{A}_{m}) +
   \ \ \ \ \ \ \ \ \ \ \ \ \ \ \ \ \ \ \ \ \ \ \ \ \
   \een
   \be
   \ \ \ \ \ \ \ \ \ \ \ \ \ \ \ \ \ \ \ \ \ \ \ \ \
   +
  \frac{2}{m^{2}} \,
    \left[  m^{2} \bar{A}_{0} + ( \partial_{m} \bar{\Pi}_{m} )
     \right] \,
    \left[  m^{2} {A}_{0} + ( \partial_{m} \bar{\Pi}_{m} )
     \right]
\label{eq:trinact}
 \ee
will prescribe the motion  equivalent to the one obtained within
approach A, on the classical level at least.

\subsubsection{Approach C \label{C}}

A key shortcoming of both the above approaches A and B (which lead
to the same equations of motion as obtained in the ``economical"
setting) is that they prove less convenient for quantization
purposes. Apparently, this would make the Dirac's theory
manifestly disqualified in comparison with the economical
approach. Fortunately, it is possible (as well as not too
difficult) to re-adapt the Dirac's procedure for quantization
purposes.

In the first step we calculate the Poisson brackets for all the
constraints and define a certain auxiliary matrix ${\cal M}$,
 \be
\left( {\cal M}^{\rule{0pt}{1ex}} (\vec{x}, \vec{y}) \right)_{J,K}
=
    \{ \theta^{(J)} (\vec{x}) , \theta^{(K)} (\vec{y}) \}_{\rm P}
 ,\ \ \ \ \ \ J,K=1, 2, 3, 4
\label{eq:ctrnact}
 \ee
where we renumbered
 \begin{eqnarray}
  \theta^{(1)} = \theta^{\rm (I)},
    &\rule{1cm}{0cm}&
    \theta^{(2)} = \theta^{\rm (II)},
    \label{eq:patnact} \\
  \theta^{(3)} = \bar{\theta}^{\rm (I)}, &&
    \theta^{(4)} = \bar{\theta}^{\rm (II)}.
    \nonumber
 \end{eqnarray}
We find that
 \be
 {\cal M}_{JK} (\vec{x}, \vec{y}) =
   {\bf M}_{JK} \, \delta^{(3)} (\vec{x} - \vec{y})
\label{eq:sedmnact}
 \ee
with the boldface symbol $\bf{M}$ denoting an antidiagonal matrix,
 \be
  \label{eq:devatenact}
  {\bf M} =
    \left(
       \begin{array}{cccc}
         0      & 0     & 0      & m^{2} \\
     0      & 0     & -m^{2} & 0     \\
     0      & m^{2} & 0      & 0     \\
     -m^{2} & 0     & 0      & 0
       \end{array}
     \right),
     \ \ \ \ \ \  {\bf M}^{-1} = - \frac{1}{m^{4}} {\bf M}.
\label{eq:osmnact}
 \ee
As long as the matrix ${\bf M}$ is non-singular and invertible we
can easily insert
 \be
  {\cal M}_{JK}^{-1} (\vec{x}, \vec{y}) =
   {\bf M}_{JK}^{-1} \, \delta^{(3)} (\vec{x} - \vec{y})
  \label{eq:dvacet}
 \ee
in an explicit definition of the Lagrange multipliers of approach
B and obtain
 \be
  \lambda^{(J)} = - \sum_{K=1}^{4} \{ H^{0} , \theta^{(K)} \}_{\rm P}
    {\bf M}_{KJ}^{-1}.
  \label{eq:dvacetjedna}
 \ee
The numbering of $\lambda$s is inherited from $\theta$s (cf.\
eq.~(\ref{eq:patnact})) and gives the final compact formula for
the Hamiltonian,
 \be
\label{eq:dvacetdva}
  {\cal H}^{\rm (B)} = {\cal H}^{0} -
    \sum_{J,K=1}^{4} \{ H^{0} , \theta^{(K)} \}_{\rm P}
    {\bf M}_{KJ}^{-1} \theta^{(J)}, \ \ \ \ \
\ \ \
  {H}^{0} = \int d \vec{x} \, {\cal H}^{0}.
\label{eq:dvacettri}
 \ee
Still, our work is not finished yet since the consistency of the
recipe still requires a replacement of the Poisson brackets by the
so called Dirac brackets,
 \be \{ X , Y \}_{\rm
 D} =
  \{ X , Y \}_{\rm P} - \sum_{J,K=1}^{4} \: \int d \vec{w} d \vec{z}
   \,
  \{ X , \theta^{(J)} (\vec{w}) \}_{\rm P} \,
    {\cal M}_{JK}^{-1} (\vec{w}, \vec{z}) \,
    \{ \theta^{(K)} (\vec{z}) , Y \}_{\rm P} .
\label{eq:dvacetctyri}
 \ee
Fortunately, one can show very easily that the equations of motion
which result from approach B are practically equivalent to those
which can be obtained from eq.~(\ref{eq:dvacetctyri}). Roughly
speaking, there emerge differences reducible to certain
combination of terms which would vanish due to the constraints.
Ref.~\cite{GT} may be consulted for details.

The main advantage of the replacement of the Poisson brackets by
their Dirac alternative is that the Dirac brackets annihilate {\em
all} the constraints,
 \be
  \{ X, \theta^{(J)} \}_{\rm D} = 0 \ \ \ \ \
     \mbox{for } J \in \{ 1, 2, 3, 4 \} \mbox{ and } \forall X.
     \label{eq:dvacetpet}
 \ee
Such a property will prove extremely useful during quantization
since the use of the Dirac brackets (in place of the Poisson ones)
enables us to work with the ``core" Hamiltonian ${\cal H}^{0}$
only.

\section{Quantization in Heisenberg picture \label{tri} }

\subsection{``Economical"  formulation \label{alfa}}

In the Heisenberg picture the quantization of the system yields a
replacement of the  generalized coordinates ${A}_m$ and $\Pi_n$
and of their complex conjugates by the non-commutative operators
which satisfy the canonical equal-time commutation relations,
  \ben
 [{A}_m({\bf x}),\ \Pi_n({\bf y})]=i \delta_{nm}\delta^{(3)}
 ({\bf x}-{\bf y}),\
 \een
 \be
 [\bar{{A}}_m({\bf x}),\ \bar{\Pi}_n({\bf y})]
 =i \delta_{nm}\delta^{(3)}({\bf x}-{\bf y}),\ \ n,m=1,2,3.
 \label{eq:zero}
 \ee
They are---up to the factor $i$---parallel to the Poisson-bracket
rules.

The time-evolution of the operators is determined by the modified
equations (\ref{pois}),
 \ben
 \dot{{A}}_k=-i[H,{A}_k]=-{\bar{\Pi}_k}+\frac{1}{m^2}\partial_k
 (\partial_n\bar{\Pi}_n),
 \een
 \ben
    \dot{\bar{\Pi}}_k=-i[H,\bar{\Pi}_k]
    =m^2{{A}}_k-\triangle{{A}}_k
    +\partial_k(\partial_n{{A}}_n),
    \een
 \ben
 \dot{\bar{{A}}}_k=-i[H,\bar{{A}}_k]=-{{\Pi}_k}
 +\frac{1}{m^2}\partial_k(\partial_n{\Pi}_n),
 \een
 \be
    \dot{\Pi}_k=-i[H,\Pi_k]=m^2\bar{{A}}_k
    -\triangle\bar{{A}}_k+\partial_k(\partial_n\bar{{A}}_n).
    \label{oo}
    \ee
An arrangement of ${A}$ and $\bar{\Pi}$ in a single six-component
set $\Psi^T=(m{\bf {A}},i\bar{\bf \Pi})$ is sometimes being used
\cite{onidva}.

\subsection{Dirac theory -- approach A \label{beta}}

In place of the six-dimensional Hamiltonians $H$ pertaining to the
preceding approach let us now switch to the so called Dirac's
approach \cite{Duirac,GT} where, as we already outlined, certain
more-dimensional Hamiltonians with redundant components are being
used.

The form of the commutators should result from their
Poisson-bracket predecessors. Nevertheless, we have to keep in
mind the existence of the additional, ``redundant" degrees of
freedom now. Thus, an additional pair of operators with the
non-trivial commutators emerges,
 \ben
 [{A}_{0}({\bf x}),\ \Pi_{0}({\bf y})]=i \delta^{(3)}
 ({\bf x}-{\bf y}),\
 \een
 \be
 [\bar{{A}}_{0}({\bf x}),\ \bar{\Pi}_{0}({\bf y})]
 =i \delta^{(3)}({\bf x}-{\bf y}).
 \label{eq:dvacetosm}
 \ee
Performing all the calculations we may collect the extended set of
equations,
 \ben
 \dot{{A}}_k=-i[H^{\rm (A)},{A}_k]=
   -{\bar{\Pi}_k} - (\partial_{m} A_{0}) ,
 \een
 \ben
    \dot{{\Pi}}_k=-i[H^{\rm (A)},{\Pi}_k]
    =m^2{\bar{A}}_k-\triangle{\bar{A}}_k
    +\partial_k(\partial_n{\bar{A}}_n) -
     (\partial_{m} \Pi_{0}),
    \een
 \ben
 \dot{{{A}}}_{0}=-i[H^{\rm (A)},{{A}}_{0}]=
   - (\partial_{m} {A}_m),
 \een
 \be
    \dot{\Pi}_{0}=-i[H^{\rm (A)},\Pi_{0}]=
    - m^2 \bar{{A}}_{0}
    -(\partial_{m} \Pi_{m}).
    \label{eq:dvacetsest}
    \ee
For the sake of brevity we do not display the complementary,
complex conjugate version of these equations.

\subsection{Dirac theory -- version B \label{gama}}

The Hamiltonian $H^{\rm (B)}$ enters  formulae
 \ben
 \dot{{A}}_k=-i[H^{\rm (B)},{A}_k]=
   -{\bar{\Pi}_k} +
     \frac{2}{m^2}\partial_k
 (\partial_n \bar{\Pi}_n) +
     (\partial_{m} A_{0}) ,
 \een
 \ben
    \dot{{\Pi}}_k=-i[H^{\rm (B)},{\Pi}_k]
    =m^2{\bar{A}}_k-\triangle{\bar{A}}_k
    +\partial_k(\partial_n{\bar{A}}_n) - (\partial_{m} \Pi_{0}),
    \een
 \ben
 \dot{{{A}}}_{0}=-i[H^{\rm (B)},{{A}}_{0}]=
   - (\partial_{m} {A}_m),
 \een
 \be
    \dot{\Pi}_{0}=-i[H^{\rm (B)},\Pi_{0}]=
    + m^2 \bar{{A}}_{0}
    + (\partial_{m} \Pi_{m}).
    \label{eq:dvacetsedm}
    \ee
In essence, the practical consequences of the procedure remain
virtually the same as before.

\subsection{Dirac theory -- approach C \label{delta}}

The major change comes with the necessity of using the
``Dirac-originating" commutators in place of the
``Poisson-originating" ones. This leads to the following two
nontrivial rules
 \ben
 [{A}_{0}({\bf x}),\ \Pi_{0}({\bf y})]= 0,\
 \een
 \be
 [{{A}}_{0}({\bf x}),\ \bar{B}_{m}({\bf y})]
 = - i \frac{\partial}{\partial y_{m}} \delta^{(3)}
 ({\bf x}-{\bf y})
 \label{eq:dvacetdevet}
 \ee
(cf.\ relations of eqs.~(\ref{eq:zero}) and (\ref{eq:dvacetosm})).
In the light of the latter ``distortion" of the commutators we are
allowed to employ the Hamiltonian $H^{0}$ and to arrive at the set
of the equations
 \ben
 \dot{{A}}_k=-i[H^{\rm (B)},{A}_k]=
   -{\bar{\Pi}_k} -
     \frac{1}{m^2}\partial_k
 (\partial_n\bar{\Pi}_n) -
     2 (\partial_{m} A_{0}) ,
 \een
 \ben
    \dot{{\Pi}}_k=-i[H^{\rm (B)},{\Pi}_k]
    =m^2{\bar{A}}_k-\triangle{\bar{A}}_k
    +\partial_k(\partial_n{\bar{A}}_n)
    \een
 \ben
 \dot{{{A}}}_{0}=-i[H^{\rm (B)},{{A}}_{0}]=
   - (\partial_{m} {A}_m),
 \een
 \be
    \dot{\Pi}_{0}=-i[H^{\rm (B)},\Pi_{0}]=0.
    \label{eq:tricet}
    \ee
We may observe and conclude that all the available recipes are
equivalent because they differ only by the terms which are
proportional to certain combinations of the constraints.

\section{Time-evolution and ${\cal PT}-$symmetry \label{styria}}

\subsection{Six-dimensional Hamiltonian \label{sixD}}

After the ``economical" quantization we arrive at the formal
Schr\"{o}dinger equation
 \be
 i\frac{\p}{\p t}\,|\,\Psi\rangle =H\,|\,\Psi \rangle\,
 \label{ham}
 \ee
which prescribes the time-evolution of the operators in our
system. The quantum Hamiltonian $H=H^{\rm (VB)}$ of the vector
boson is a matrix
 \be
 H=\left(\begin{array}{cc}0&-m+\frac{\grad\div}{m}\\-m
 +\frac{\nabla^2}{m}-
 \frac{\grad\div}{m}&0\end{array}\right).
 \label{brouk}
 \ee
(cf. \cite{onidva}). In the coordinate representation it has the
$6\times 6-$dimensional and manifestly non-Hermitian partitioned
matrix structure,
 \ben
 H \sim \left(\begin{array}{cccccc}0&0&
 0&\partial_1^2-m^2&\partial_1\partial_2&\partial_1\partial_3\\
 0
 &0&0
 &\partial_2\partial_1&\partial_2^2-m^2&\partial_2\partial_3\\
 0&0&0&\partial_3\partial_1&\partial_3\partial_2
  &\partial_3^2-m^2\\
 -\partial_1^2-\omega^2&-\partial_1\partial_2
 &-\partial_1\partial_3&0&0&0\\
 -\partial_2\partial_1&-\partial_2^2-\omega^2
 &-\partial_2\partial_3&0&0&0\\
 -\partial_3\partial_1&-\partial_3\partial_2
 &-\partial_3^2-\omega^2&0&0&0\end{array}\right).
 \een
As a ${\cal P}-$pseudo-Hermitian operator,
 \ben
 H^{\dagger}={\cal P} H{\cal P}^{-1} \neq H,\ \ \ \ \ \ \ \ \ \ \
  {\cal P}=\left(\begin{array}{cc}0&I_3\\I_3&0
  \end{array}\right)
  \een
it allows a quantization in a  finite box of a side-length $L$.
Using the periodic boundary conditions imposed at its walls one
arrives at the discrete bound-state energies on the mass shell,
 \ben
 p_{0,{\bf n}}=\pm\sqrt{{\bf p}_{\bf n}^2+m^2},\ \ \ \ \ \
 \ p_{j,{\bf n}}=\frac{2\pi}{L}{\bf n}
 \een
with ${\bf n}$ being a vector of integer components. Vice versa,
the continuity of the energies can be restored in the $L \to
\infty $ limit whenever necessary.

At a finite $L$ the six independent eigenvectors with eigenvalues
$\pm p_0$ may be most easily constructed in terms of the
above-introduced functions in momentum representation,
 \ben
 \Psi^{(\pm)}_k(p,j)=\left\{\begin{array}{ll}
 m{A}^{(\pm)}_k(p,j),& k=1,2,3\\
   \mp \omega{A}^{(\pm)}_k(p,j)+p_k{A}^{(\pm)}_0(p,j)
             ,& k=4,5,6\end{array}\right.
             , \ \ \ \ \ \ \ j=1,2,3.
             \een
Here we abbreviated
 \ben
 {A}^{(\pm)}({{p}},1)
 =\frac{{A}_1^{(\pm)}(p)+i{A}_2^{(\pm)}(p)}{\sqrt{2}},
 \een
 \ben
 {A}^{(\pm)}({{p}},2)
 =\frac{{A}_1^{(\pm)}(p)-i{A}_2^{(\pm)}(p)}{\sqrt{2}},
 \een
 \ben
 {A}^{(\pm)}({{p}},3)={A}_3^{(\pm)}(p)
 \een
with  three linearly independent  auxiliary amplitudes
${A}_i^{(\pm)}(p)$ introduced originally as perpendicular to the
momentum four-vector,
 \be
 {A}^{(\pm)}_1({\bf p})=(0,{\bf e}_1),\ \ {A}^{(\pm)}_2({\bf p})
 =(0,{\bf e}_2),
 \ \
 {A}^{(\pm)}_3=\left(\frac{|{\bf p}|}{p_0},
 \frac{{\bf p}}{|{\bf p}|}\right).
 \label{pes}
 \ee
The  unit vectors ${\bf e}_1$, ${\bf e}_2$ are chosen as mutually
perpendicular, $({\bf e}_1,{\bf\bar{e}}_2)=0$, and perpendicular
also to the momentum vector ${\bf p}$,
 \ben
 {\bf {A}}^{(\pm)}({{\bf p}})=a_1^{(\pm)}({\bf p}){\bf e}_1
 +a_2^{(\pm)}({\bf p}){\bf e}_2
 +a^{(\pm)}_3({\bf p})\frac{{\bf p}}{|{\bf p}|}.
 \een
In order to make our solutions (confined in a box!) less ambiguous
we introduce another observable which commutes with $H$,
 \ben
 {\Lambda}_{i,j}({\bf p})={\Lambda}_{i+3,j
 +3}({\bf p})=-\frac{1}{2}+
 \frac{\frac{3}{2}{p}_i{p}_j+i|{\bf p}|
 \varepsilon_{ijk}{p}_k}{{\bf p}^2},
 \een
 \be
  \hat{\Lambda}_{i,j+3}({\bf p})=\hat{\Lambda}_{i+3,j}
 ({\bf p})=0,\ \ i,j=1,2,3.
 \label{27}
 \ee
This is a block diagonal operator which distinguishes the spin
projection into momentum direction. In terms of physics it
corresponds to the helicity of our vector bosons. It is easy to
verify that the latter operator is also pseudo-Hermitian with
respect to ${\cal P}$,
 \ben
 \Lambda^{\dagger}={\cal P}\Lambda{\cal P}^{-1} .
 \een
We obtain a complete right-action Schr\"{o}dinger equation for the
fixed-helicity states,
 \ben
  H|\pm \omega,h\rangle=\pm\omega|\pm \omega,h\rangle,
  \ \ \hat{\Lambda}|\omega ,
 h\rangle=h|\omega,h\rangle,\
 \ h=\pm1,\ 0\label{22bi}
 \een
as well as the similar relation for the action of our observables
to the left,
 \ben
 \ll \pm \omega,h| H=\pm\omega \, \ll \pm \omega,h|,\ \
 \ \ \ \ \ \ll \pm \omega,h|\hat{\Lambda}=h\,
  \ll \pm \omega,h|,\ \ \ \
 \ h=\pm 1,\ 0.
 \label{22bc}
 \een
In a way described more thoroughly in ref. \cite{coupledchannels}
the pseudo-Hermicity of both our observables can immediately be
used to express the double-ket-marked eigenvectors of the
conjugate-operator pair of $H^{\dagger}$ and $\Lambda^{\dagger}$
in terms of the single-ket-marked eigenvectors of the pair of
operators $H$ and $\Lambda$,
 \be
 |\pm \omega,{h}\rr={\cal P}|\pm
 \omega,{h}\rangle \varrho_{p_0,h}\,.
 \label{kocka}
 \ee
One can normalize the vectors $| p_0,{h}\rangle$ and
$|p_0',{h}'\rr$ in such a way that they form a biorthonormal set,
 \ben
  \ll
 p_0,{h}|p_0',{h}'\rangle=\delta({p_0-p_0'})
 \delta_{{h}{h}'}
 \label{24kj}
 \een
(cf. ref. \cite{coupledchannels} for more details).

\subsection{Alternative eight-dimensional Hamiltonians \label{eightD}}

In the Dirac case we can write Hamiltonians in the partitioned
matrix form, too. All such matrices will be eight-dimensional.

In approach A, the matrix form of the Hamiltonian is compatible
with the related equations of motion (\ref{eq:dvacetsest}),
 \be
 H^{\rm (A)} =
   \left(
     \begin{array}{cccc}
       0  & -m + \frac{2\grad\div}{m} & \grad & 0 \\
       -m
  + \frac{\nabla^2}{m}-
 \frac{\grad\div}{m}
             & 0     & 0  & \grad \\
       -\div & 0     & 0  & 0 \\
       0     & -\div & -m & 0
     \end{array}
   \right).
 \label{tricetjedna}
 \ee
It is amusing to notice that the approach B will give the matrix
Hamiltonian in a slightly different form,
 \be
 H^{\rm (A)} =
   \left(
     \begin{array}{cccc}
       0  & -m & - \grad & 0 \\
       -m
  + \frac{\nabla^2}{m}-
 \frac{\grad\div}{m}
             & 0     & 0  & \grad \\
       -\div & 0     & 0  & 0 \\
       0     & \div & m & 0
     \end{array}
   \right)
 \label{tricetdva}
 \ee
(cf.\ equations of motion (\ref{eq:dvacetsedm})). Still another
version of the Hamiltonian will emerge from the approach C,
\be
 H^{0} =
   \left(
     \begin{array}{cccc}
       0  & -m - \frac{\grad\div}{m} & - 2\grad & 0 \\
       -m
  + \frac{\nabla^2}{m}-
 \frac{\grad\div}{m}
             & 0     & 0  & 0 \\
       -\div & 0     & 0  & 0 \\
       0     & 0     & 0  & 0
     \end{array}
   \right)
 \label{tricettri}
 \ee
(cf.\ equations of motion (\ref{eq:tricet})). All the latter
matrices are non-Hermitian. However, they are also ${\cal
PT}-$symmetric in the way which employs the {\em same}, constant
and partitioned matrix operator
 \be
 {\cal P} =
   \left(
     \begin{array}{cccc}
       0     & I_{3 }& 0   & 0  \\
       I_{3} & 0     & 0   & 0  \\
       0     & 0     & 0   & -1 \\
       0     & 0     & -1  & 0
     \end{array}
   \right),
 \label{tricetctyri}
 \ee
in all the three contexts. It is worth emphasizing that the use of
the Hamiltonian pertaining to the Dirac approach C seems
definitely preferable from the point of view of  quantization.

\section{Summary\label{sest}}

Once we assume that the pair of a quantum Hamiltonian $H$ and of
some auxiliary ``observable" $\Lambda$ are independent and
diagonalizable {\em non-Hermitian} operators with discrete
spectra, we may study this pair as a special mathematical
realization of a ${\cal PT}-$symmetric quantum model
\cite{coupledchannels}. In our present paper we paid particular
attention to the related necessity of the explicit knowledge of a
{\em sufficiently simple} auxiliary generalized parity operator
${\cal P}$. In this sense we have shown here that the classical
Proca's field offers a new and {\em feasible} example of a ${\cal
PT}-$symmetric system, this time with an immediate physical
vector-boson interpretation.

\subsection{Proca system with pseudo-Hermitian observables}

Let us summarize our preceding text as a constructive presentation
of the triplet of operators $H$, $\Lambda$ and ${\cal P}$ on a
Hilbert space  ${\cal V}$ which fulfills the relations
 \be
  H^{\dagger}={\cal P} H {\cal P}^{-1},\ \ \Lambda^{\dagger}
 ={\cal P} \Lambda {\cal P}^{-1}
  \label{19,5}
  \ee
where ${\cal P}$ is an extremely simple though nontrivial and
positively indefinite Hermitian automorphism on ${\cal V}$.

In the related literature, the ${\cal PT}-$symmetric operators $H$
and $\Lambda$ are also often called pseudo-Hermitian or ${\cal
P}-$pseudo-Hermitian. One of the most important consequences of
their ${\cal PT}-$symmetry is well known to lie in the facilitated
possibility of the construction of the physical or
``representation" Hilbert space and metric using the factorization
ansatz $\Theta = {\cal CP}$ where ${\cal C}$ may be called a
``charge" \cite{BBJo}.

Let us repeat that having assumed the simultaneous
diagonalizability {\em and} a non-degeneracy of the (real or
complex) spectra of {\em both} $H$ and $\Lambda$ we can, at least
in principle, find their mutually bi-orthogonal left and right
eigenvectors,
 \be
 H|n,{h}\rangle=E_n|n,{h}\rangle,
 \ \ \ \ \ \
  \ll m,{h}|H=\ll
 m,{h}|E_m,\
 \ee
 \be
   \Lambda|n,{h}\rangle={h}|n,{h}\rangle, \ \ \ \ \
  \ll m,{h}|\Lambda=\ll
 m,{h}|{h},
 \ee
as well as their formal spectral representation,
 \be
 H=\sum_n\,|n,{h}\rangle \frac{E_n}
 {\langle\langle n,{h}|n,{h}\rangle}\,\langle\langle n,{h}|,
 \ee
 \be
 \Lambda=\sum_{h}\,|n,{h}\rangle \frac{{h}}
 {\langle\langle n,{h}|n,{h}\rangle}\,\langle\langle n,
 {h}|
  .\label{19,7}
  \ee
Due to (\ref{19,5}), the left and right eigenvectors are mutually
related by ${\cal P}$
 \bea
 {\cal P}|n,{h}\rangle\sim\rho_{n,{h}}|n,{h}\rr,
 \label{kos}
 \eea
where the ``quasi-parity" $\rho_{n,{h}}$ is any real or complex
overlap between the eigenvectors before their
biorthonormalization,
 \ben
 \ll n,{h}|n',{h}'\rangle=\delta_{n,n'}\delta_{{h},
 {h}'}\rho_{n,{h}}.
 \een
{\it A priori}, the spectra of the eigenvalues of the operators
$H$ and/or $\Lambda$ which satisfy the relations (\ref{19,5}) may
be either purely real or containing certain conjugate pairs of
eigenvalues or containing certain ``exceptional" points where the
diagonalizability has been lost \cite{Heiss}.

Whenever one contemplates the former option only, the reality of
the eigenvalues opens a way towards a physical interpretation of
the model where the eigenvalues of operators $H$ and $\Lambda$ may
be observable and where the distribution of these eigenvalues can
be experimentally measured. In the literature, a successful
attempt of such a type has recently been completed for the
spinless Klein-Gordon system \cite{Ali}. In our present paper we
made the first steps in the similar direction for spin one. For
several alternative versions of the Hamiltonian $H$ and associated
helicity operator $\Lambda$ we constructed certain ``sufficiently
simple" operators ${\cal P}$ playing the role of an indefinite
metric which characterizes the relativistic Proca field of the
massive vector bosons with spin one.

\subsection{Outlook}

Our present study has been inspired by the well known fact that in
the massless limit of the vector bosons the construction of the
physical metric becomes trivial. In particular, the operator
$\Theta$ of eq. (\ref{QH}) degenerates to a projector in a way
explained by Gupta and Bleuler \cite{GB} (cf. also Appendix A
below). In this sense our present study could be perceived as a
preparatory clarification of several ${\cal PT}-$symmetry-related
questions emerging during the canonical quantization of the
massive model.

Even in the absence of interactions we may expect that the
construction of the positively definite $\Theta$ will be much less
easy than in the electromagnetic massless case. Its details seem
to represent an open problem up to now \cite{Nikitin}. At the same
time, the construction of $\Theta$ should be considered urgent at
any spin. Its phenomenological relevance stems from the necessity
of a search for a compromise between the complexity of the
field-theoretical calculations and the comparative simplicity of
the various pragmatic semi- or non-relativistic models of the
bound states (say, in a pionic atom, etc). We believe that the
challenge of the construction of $\Theta$ for spin one
\cite{onidva} might also re-open several new purely theoretical
questions paralleling the ones encountered already in the
zero-spin context of the Feshbach-Villars version of the
first-quantized Klein-Gordon equation \cite{Ali}.

\section*{Acknowledgements}

Participation of VJ and MZ supported by IRP AV0Z10480505 and by
the ``Doppler Institute" MSYS project Nr. LC06002.

\newpage

\section*{Appendix A. Parallels between massive spin-one bosons
and massless photons}

In a way resembling the description of the massless (i.e.,
electromagnetic) fields let us now change the notation and
introduce an antisymmetric tensor $F^{\mu\nu}$ such that
 \bea
 F^{ij}&=&-\varepsilon_{ijm}B_m,\nonumber\\
      F^{0j}&=&-E_j,\ \ i,j=1,2,3.
      \nonumber
      \eea
In a fixed inertial frame equations (\ref{1}) then acquire the
form
 \be
 \begin{array}{cclccl}{\bf {B}}&
 =&\rot {\bf {A}},&\f{\partial{\bf A}}{\partial t}&
 =&-{\bf {E}}-\grad A_{0}\\
 A_0&=&-m^{-2}\div{\bf {E}},&\f{\partial{\bf {E}}}
 {\partial t}&=&m^2{\bf {A}}+\rot {\bf {B}}\,.\end{array}
 \label{12}
 \ee
An elimination of ${\bf B}$ and $A_0$ gives
 \bea
 \f{\partial{\bf {A}}}{\partial t}=-{\bf {E}}
 +\f{\nabla^2{\bf {E}}}{2m^2}
 +\f{\hat{q}}{2m^2}{\bf {E}},\ \
 \f{\partial{\bf {E}}}{\partial t}=m^2{\bf {A}}
 -\f{\nabla^2{\bf {A}}}{2}
 +\f{\hat{q}}{2}{\bf {A}},
 \label{13}
 \eea
where we abbreviated
 \ben
 \hat{q}=2\grad\div-\nabla^2\,.
 \een
Finally, another abbreviation
 \bea
 \Phi^T=({\bf{u}},{\bf v}),\ \ {\bf u}={\bf {E}}+im{\bf {A}},\
  \ {\bf v}={\bf {E}}-im{\bf {A}}
  \label{14}
  \eea
converts equations (\ref{13}) into the  six-component evolution
rule
 \ben
  i\f{\partial{\Phi}}{\partial
 t}=\left(\begin{array}{cc}m-\f{\nabla^2}{2m}&-\f{\hat{q}}{2m}
 \\
 \f{\hat{q}}{2m}&-m+\f{\nabla^2}{2m}
 \end{array}\right)
 \Phi=
 \ \ \ \ \ \ \ \ \ \ \ \ \ \ \ \ \ \ \ \ \ \ \ \ \ \
 \een
 \be
 \ \ \ \ \ \ \ \ \ \ \ \ \ \ \ \ \ \ \ \ \ \ \ \ \ \
 =\left(\begin{array}{cc}m-\f{\nabla^2}{2m}&-\f{\nabla^2}{2m}
 +\f{({\bf S}{\bf\nabla})^2}{m}\\
 \f{\nabla^2}{2m}-\f{({\bf S}{\bf\nabla})^2}{m}&
 -m+\f{\nabla^2}{2m}
 \end{array}\right)\Phi=H_{TS}\Phi
 .\label{15}
 \ee
The matrix operator $H_{TS}$ is not Hermitian and fulfills the
relation
 \be
  H_{TS}^{\dagger}=\sigma_3 H_{TS}\,\sigma_3, \ \ \ \ \
 \sigma_3=\left(\begin{array}{cc}I_3&0\\0&-I_3
 \end{array}\right)
 \label{19}
 \ee
shared by the spinless and spin-one cases \cite{FV,case}.

\end{document}